\documentclass[aps,prb,amsmath,amssymb,footinbib,showpacs,twocolumn]{revtex4-2}
\usepackage{amsmath}
\usepackage{amssymb}
\usepackage{amsthm}
\usepackage{setspace}
\usepackage{graphicx}
\usepackage{braket}
\usepackage{mathrsfs}
\usepackage{float}
\usepackage[colorlinks = true,linkcolor = blue,urlcolor  = blue,citecolor = blue,anchorcolor = blue]{hyperref}
\usepackage[utf8]{inputenc}
\usepackage[english]{babel}
\usepackage{bm}

\begin{document}

\title{Anomalous transport in pseudospin-1 fermions}

\author{Adesh Singh}
\affiliation{School of Physical Sciences, Indian Institute of Technology Mandi, Mandi 175005, India}
\author{G. Sharma}
\affiliation{School of Physical Sciences, Indian Institute of Technology Mandi, Mandi 175005, India}

\begin{abstract}
Electronic transport in the $\alpha-\mathcal{T}_3$ model of pseudospin-1 fermions with a finite gap is studied within the semiclassical Boltzmann approximation. We show that coupling of the orbital magnetic moment to the external magnetic field, which is otherwise absent in the massless model, breaks valley symmetry, results in finite and measurable corrections to the longitudinal and Hall conductivity, and yields anomalous Hall conductivity due to the Berry curvature. We also show that, remarkably, magnetoresistance induced by the orbital magnetic moment can be either positive or negative; the sign depends on the amount of disorder, and is different for both conventional and anomalous contributions to the magnetoresistance. Recent material advances and upcoming experiments on cold atoms that may realize pseudospin-1 fermions makes our study timely and appropriate. 
\end{abstract}

\maketitle
\section{Introduction}
The importance of Berry's phase and Berry curvature effects in solids is now very well established~\cite{xiao2010berry, sundaram1999wave,chang1995berry,chang1996berry,chang2008berry}. While Berry curvature in solids leads to anomalous Hall effects~\cite{nagaosa2010anomalous}, more recently, it has been realized that it can also explain exotic effects, such as the chiral anomaly in Weyl semimetals, within the semiclassical Boltzmann approximation~\cite{son2013chiral}. 
While Berry curvature has received significant attention, the role of the anomalous orbital magnetic moment has been explored far less. Cognate to the Berry curvature, the orbital magnetic moment also owes its origin to the geometrical effects in the Bloch bands~\cite{chang1996berry}. 
Specifically, the self-rotating angular momentum of the Bloch wave packet gives rise to an anomalous orbital magnetic moment that couples to the external magnetic field via a Zeeman-like interaction. This phenomenon has recently yielded unconventional effects in semiclassical electronic transport.~\cite{ma2015chiral,varjas2016dynamical,zhou2019valley,knoll2020negative,sharma2020sign, das2021intrinsic}.

In recent times, graphene and similar other Dirac materials have received the attention of diverse audiences due to their unconventional energy dispersion and non-trivial topology of their Bloch wavefunctions~\cite{neto2009electronic,sarma2011electronic,armitage2018weyl}. On the other hand, the interest in dispersionless flat bands has been rejuvenated after the discovery of superconductivity in twisted bilayer graphene~\cite{cao2018unconventional}. 
First proposed in Ref.~\cite{raoux2014dia}, the $\alpha-\mathcal{T}_3$ model, synthesizes flat and dispersive Dirac bands in a single model. The tight-binding description of the $\alpha-\mathcal{T}_3$ model comprises a hexagonal lattice with atoms situated at the vertices of the hexagons and their centers. Thus, the unit cell consists of three atoms, constructing the electronic-state description that of a triple-component fermion (pseudospin-1 fermions). Varying the parameter $\alpha$ in the $\alpha-\mathcal{T}_3$ lattice, one interpolates between graphene ($\alpha=0$) and the dice lattice ($\alpha=1$). When $\alpha\in(0,1]$, the band structure is remarkably independent of $\alpha$ consisting of a flat band at zero energy intersecting a Dirac cone. Experimentally, the $\alpha-\mathcal{T}_3$ model can be realized in cold atoms, Josephson arrays, trilayers of cubic lattices, and Hg$_{1-x}$Cd$_x$Te quantum wells~\cite{wang2011nearly, rizzi2006phase, malcolm2015magneto, serret2002vortex}.  

While the $\alpha-\mathcal{T}_3$ spectrum is gapped in an external magnetic field, the flat band remains primarily unaffected~\cite{raoux2014dia}. Some other recent works have discussed the stability of the flat bands to impurities, the presence of boundaries, and perturbative radiation~\cite{dey2018photoinduced,bilitewski2018disordered,oriekhov2018electronic}. In the $\alpha-\mathcal{T}_3$ model, while the Dirac cone may gap out due to these factors, simply shifting the chemical potential in one or more of the sublattices can also produce a Dirac gap (massive pseudospin-1 fermions). The massless $\alpha-\mathcal{T}_3$ model's optical, magnetic, and electronic transport properties have been studied~\cite{islam2017valley,illes2015hall, kovacs2017frequency,biswas2016magnetotransport,xu2017unconventional,illes2016magnetic}; investigations on the gapped model remain less explored. While owing certain similarities to the physics of pseudospin-1/2 fermions in massive graphene models~\cite{xiao2007valley,zhou2019valley}, as we show later, the presence of a dispersionless (or weakly dispersive) flat band in pseudospin-1 fermions strikingly changes the topology of the wavefunctions, and, more so, the geometrical quantities such as the Berry curvature and the orbital magnetic moment. The variation of these quantities with the lattice parameters $\alpha$ is insightful as well, and reflects the $\alpha\rightarrow 1/\alpha$ duality. 

Here we study the electronic transport properties of the massive $\alpha-\mathcal{T}_3$ model, focusing particularly on the nontrivial role of the anomalous orbital magnetic moment. We first discuss the implications of the Dirac gap on conventional transport in the absence and presence of an external magnetic field, specifically evaluating the longitudinal and the Hall conductivity for three different disorder profiles: $\delta-$correlated impurities, Gaussian impurities, and screened Coulomb impurities. A crucial distinction between massless and massive pseudospin-1 fermions is the presence of Berry curvature and orbital magnetic moment. While the Berry curvature and orbital magnetic moment are absent in the massless case, as we show, both acquire finite contributions in the massive model. We show that coupling the orbital magnetic moment to the external magnetic field breaks valley symmetry, results in finite and measurable corrections to the longitudinal and Hall conductivity, and yields anomalous Hall conductivity due to the Berry curvature. Strikingly, the orbital magnetic moment induces a finite magnetoresistance (longitudinal), which can be either positive or negative, different from the claim in Ref.~\cite{zhou2019valley} in the context of pseudospin-1/2 fermions. The sign of the magnetoresistance depends on the amount of disorder, switching from positive to negative for the case of conventional contribution to the Hall effect and switching vice-versa for anomalous contribution to the Hall effect. Introducing a generic model of magnetoresistance, we conjecture that both positive and negative magnetoresistance could be observed in  Ref.~\cite{zhou2019valley}.  Our analysis assumes the semiclassical Boltzmann approximation that remains valid for weak external electric and magnetic fields.

\begin{figure*}
    \centering
    \includegraphics[width=2\columnwidth]{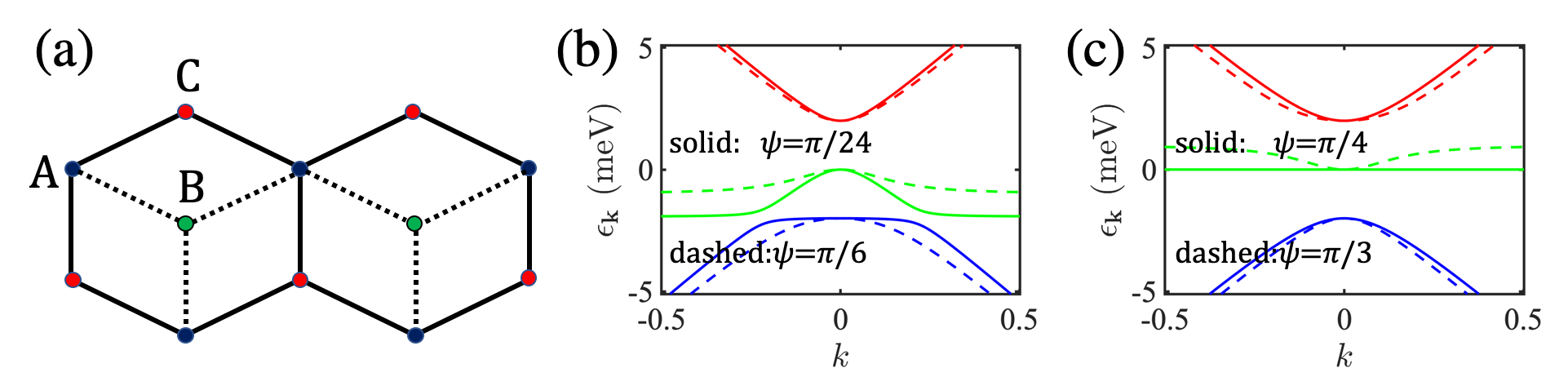}
    \caption{(a) Lattice structure of the $\alpha-\mathcal{T}_3$ model. (b) Bandstructure of the massive $\alpha-\mathcal{T}_3$ model for $\psi=\pi/24$ (solid lines), and for $\psi=\pi/6$ (dashed lines). (c) Bandstructure for $\psi=\pi/4$ (solid lines), and for $\psi=\pi/3$ (dashed lines). The dimensions of $k$ are normalized by $1/a$, where $a=50$nm. We choose $m=2$meV.}
    \label{fig:bs_no_omm}
\end{figure*}

\section{Massive pseudospin-1 fermions}
\subsection{Hamiltonian and bandstructure}
The  $\alpha-T_3$ model consists of three sublattices that we name $A$, $B$, and $C$. Sites $A$ and $C$ occupy the positions corresponding to the two sublattices in graphene, while the centre of each hexagon has an additional site, $B$. The hopping parameters among various sites are $t_{AC}=t$ and $t_{BC}=\alpha t$. As the parameter $\alpha$ varies from zero to one, the lattice structure interpolates from  graphene to a dice lattice. Fig.~\ref{fig:bs_no_omm} (a) plots the representative real-space lattice structure. 
Here, we consider the $\alpha-T_3$ model with an additional mass term $m$, such that the momentum-space Hamiltonian near the $\mathbf{K}$ point is expressed as $H_\mathbf{k}^\tau=H_0^\tau+H_m$, where~\cite{gorbar2019electron,romhanyi2015hall,ye2020quantum}
\begin{align}
    H_0^\tau &= \tau \hbar v k\begin{pmatrix}
    0 & \cos\psi e^{-i\tau\phi} & 0\\
    \cos\psi e^{i\tau\phi} & 0 & \sin\psi e^{-i\tau\phi}\\
    0 & \sin\psi e^{i\tau\phi} & 0\\
    \end{pmatrix}\nonumber\\
    H_m &= m\begin{pmatrix}
    1 & 0 & 0\\
    0 & 0 & 0\\
    0 & 0 & -1\\
    \end{pmatrix}
    \label{Eq_H}
\end{align}
Here $\tau=\pm 1$ stands for the valley index, $v$ is the velocity parameter, $\tan\psi= \alpha$, $\tan\phi=k_y/k_x$, and $m$ is assumed to have the dimensions of energy. When $m=0$, the energy spectrum is independent of $\alpha$ and the eigenvalues are $\epsilon_\mathbf{k} = 0$, $+\hbar v_F k$, and $-\hbar v_F k$, corresponding to a flat zero-energy band intersecting the linearly dispersing Dirac cone. While disorder or magnetic field may generate a nonzero mass $m$, it may also be generated by shifting the chemical potential on sublattices $A$ and $B$ or by making the atoms non-identical on the sublattices. In the presence of the mass term, the spectrum is solved by solving the following cubic equation: 
\begin{align}
    \epsilon_\mathbf{k}^3 - \epsilon_\mathbf{k} (m^2 + \hbar^2v^2k^2) - m\hbar^2 v^2 k^2 \cos(2\psi) = 0
\end{align}

The above cubic equation may be solved via Carnado's method. The exact analytical expressions for the band dispersions for arbitrary values of $m$ and $\psi$ could be more illuminating, so we do not discuss them here; below, we present their approximate form in some exceptional limiting cases.
\begin{align}
      \epsilon_\mathbf{k} (\psi=\pi/4)&=\left\{
    \begin{array}{l}
      +\sqrt{\hbar^2 v^2 k^2 + m^2}\\
      0 \\
      -\sqrt{\hbar^2 v^2 k^2 + m^2}
    \end{array}
  \right.\\
    \epsilon_\mathbf{k} (\hbar v k\gg m)&=\left\{
    \begin{array}{l}
      \tau\hbar v k+ m  \cos (2 \psi )/2+\mathcal{O}\left(m^2\right)\\
      -m\cos(2\psi)+\mathcal{O}\left(m^2\right) \\
      -\tau\hbar v k+ m  \cos (2 \psi )/2+\mathcal{O}\left(m^2\right)
    \end{array}
  \right.\\
    \epsilon_\mathbf{k} (\hbar v k\ll m)&=\left\{
    \begin{array}{l}
      m+m^{-1}\hbar^2v^2k^2\cos^2\psi +\mathcal{O}\left(k^3\right)\\
      -m^{-1}\hbar^2v^2k^2\cos(2\psi)+\mathcal{O}\left(k^3\right) \\
      -m-m^{-1}\hbar^2v^2k^2\sin^2\psi +\mathcal{O}\left(k^3\right)
    \end{array}
  \right.
\end{align}
When $m\neq 0$, the linearly dispersing cone gaps out and the qualitative behavior of the middle band (which is dispersionless for $m=0$) depends on the value of $\psi$, roughly scaling as $\sim-\cos(2\psi)$. Therefore, 
when $\psi<\pi/4$, the middle band hosts a hole-like pocket, and when $\psi>\pi/4$, it hosts en electron-like pocket. When $\psi=\pi/4$, the band is non-dispersive. When $k\ll m/\hbar v$, the middle and the upper bands are quadratic in $k$, and are linear when $k\gg m/\hbar v$. 
Fig.~\ref{fig:bs_no_omm} (b) and (c) present the bandstructure for a few different values $\psi$ for a fixed $m$. Since the band dispersion is isotropic with respect to the polar angle $\phi$, we plot the bandstructure as a function of radial coordinate $k$.
\subsection{Conventional magnetotransport}
We first discuss conventional magnetotransport in pseudospin-1 fermions. To this end, we focus on three specific quantities: longitudinal conductance ($\sigma_{xx}$), conventional Hall conductance ($\sigma_{xy}$), and the anomalous Hall conductance ($\sigma_{xy}^\mathrm{a}$). These are given by the following expressions in the weak-field Boltzmann approximation:
\begin{align}
    \sigma_{xx} &= \frac{e^2}{\hbar}\int{\frac{d^2k}{(2\pi)^2} \left(\frac{-\partial f_0(\epsilon_\mathbf{k})}{\partial \epsilon_\mathbf{k}}\right) \tau_\mathbf{k}\left(\frac{\partial \epsilon_\mathbf{k}}{\hbar\hspace{1mm}\partial k_x}\right)^2 },\nonumber\\
    \sigma_{xy} &= \frac{e^3\hspace{1mm}b}{\hbar}\int{\frac{d^2k}{(2\pi)^2}  \left(\frac{-\partial f_0(\epsilon_\mathbf{k})}{\partial \epsilon_\mathbf{k}}\right) \left(\frac{\tau_\mathbf{k}\hspace{1mm}\partial \epsilon_\mathbf{k}}{\hbar\hspace{1mm}\partial k_y}\right)}\nonumber\\
    &\times{\left(\frac{\partial \epsilon_\mathbf{k}}{\hbar\hspace{1mm}\partial k_y}\frac{\partial}{\partial k_x} - \frac{\partial \epsilon_\mathbf{k}}{\hbar\hspace{1mm}\partial k_x}\frac{\partial}{\partial k_y}\right)\left(\frac{\tau_\mathbf{k}\hspace{1mm}\partial \epsilon_\mathbf{k}}{\hbar\hspace{1mm}\partial k_x}\right) }, \nonumber \\
    \sigma_{xy}^\mathrm{a} &= -\int{\frac{d^2k}{(2\pi)^2} \Omega^{z}(\mathbf{k}) f_0(\epsilon_\mathbf{k})}. 
    \label{Eq_sigmas}
\end{align}
Here $b$ is the applied magnetic field, $-e$ is the charge of an electron, $f_0(\epsilon_\mathbf{k})$ is the Fermi-Dirac distribution function, $\Omega^{z}(\mathbf{k})$ is the $z-$component of the Berry curvature, and $\tau_\mathbf{k}$ is the relaxation time. The relaxation time is evaluated as~\cite{bruus2004many}:
\begin{align}
\frac{1}{\tau_\mathbf{k}}=\int \frac{d^2 k^{\prime}}{(2 \pi)^2} w_{\mathbf{k}, \mathbf{k}^{\prime}}\left(1-\cos \phi_{\mathbf{k} \mathbf{k}^{\prime}}\right),
\label{Eq_tau}
\end{align}
where  $\cos \phi_{\mathbf{k} \mathbf{k}^{\prime}}$ is the angle between the vectors $\mathbf{k}$ and $\mathbf{k}'$, $w_{\mathbf{k}, \mathbf{k}^{\prime}}$ is the scattering amplitude between the states $|\mathbf{k}\rangle$ and $|\mathbf{k}'\rangle$, which strongly depends on the type of impurity scattering considered. Although anisotropy on the Fermi surface invalidates the use of Eq.~\ref{Eq_tau}, this is irrelevant here since the band dispersion is isotropic. Here we study the effects of three different types of impurities: (i) $\delta-$correlated impurities, (ii) screened Coulomb impurities, and (iii) Gaussian impurities. The general form of the impurity potential is expressed as a sum of impurity potentials located at sites $\mathbf{R}_i$:
\begin{align}
    V_\mathrm{imp}(\mathbf{r}) = \sum\limits_i u(\mathbf{r}-\mathbf{R}_i),
\end{align}
where the function $u(\mathbf{r})$ takes the following form for each of the impurity types: 
\begin{align}
    u^\delta(\mathbf{r}) &= V_\delta\delta(\mathbf{r}),\nonumber\\
    u^c(\mathbf{r}) &= \frac{e^2e^{-rk_\mathrm{TF}}}{4\pi\epsilon_0\kappa r},\nonumber\\
    u^g(\mathbf{r}) &= V_g e^{-r^2/2\sigma^2}.
\end{align}
The superscripts $c$, $\delta$, and $g$ stand for Coulomb, $\delta$ (delta), and Gaussian impurities respectively. The various parameters used above are explained as follows: (i) for $\delta-$disorder: $V_\delta$ the bare impurity strength, (ii) for Coulomb disorder: $\kappa$ is the dielectric constant, and $k_\mathrm{TF}$ is the Thomas-Fermi screening wavevector, (iii) for Gaussian disorder: $V_g$ the bare impurity strength, and $\sigma$ is the standard deviation of the Gaussian function. 
The scattering rate for each of the impurity type can be evaluated in the lowest Born approximation to be~\cite{bruus2004many}:
\begin{align}
w^\delta_{\mathbf{k}, \mathbf{k}^{\prime}}&=\frac{2 \pi}{\hbar} \frac{n^\delta_i}{\mathcal{A}} V_\delta^2\left|\left\langle\mathbf{k}^{\prime}| \mathbf{k}\right\rangle\right|^2 \delta\left(\epsilon_{\mathbf{k}}-\epsilon_{\mathbf{k}^{\prime}}\right),\nonumber\\
w^c_{\mathbf{k}, \mathbf{k}^{\prime}}&=\frac{2 \pi}{\hbar} \frac{n^c_i}{\mathcal{A}} \left(\frac{e^2}{2\epsilon_0\kappa (|\mathbf{k}-\mathbf{k}'|+k_\mathrm{TF})}\right)^2\left|\left\langle\mathbf{k}^{\prime}|\mathbf{k}\right\rangle\right|^2 \delta\left(\epsilon_{\mathbf{k}}-\epsilon_{\mathbf{k}^{\prime}}\right),\nonumber\\
w^g_{\mathbf{k}, \mathbf{k}^{\prime}}&=\frac{2 \pi}{\hbar} \frac{n^\delta_g}{\mathcal{A}} V_g^2e^{-q^2\sigma^2}\sigma^4\left|\left\langle\mathbf{k}^{\prime}| \mathbf{k}\right\rangle\right|^2 \delta\left(\epsilon_{\mathbf{k}}-\epsilon_{\mathbf{k}^{\prime}}\right),
\end{align}
where $\mathcal{A}$ is the area, $|\mathbf{k}\rangle$ is the energy eigenfunction obtained by diagonalizing Eq.~\ref{Eq_H}, and $\{n_i^\delta, n_i^c, n_i^g\}$ represent the impurity density for different impurity types. 
\begin{figure}
    \centering
    \includegraphics[width=\columnwidth]{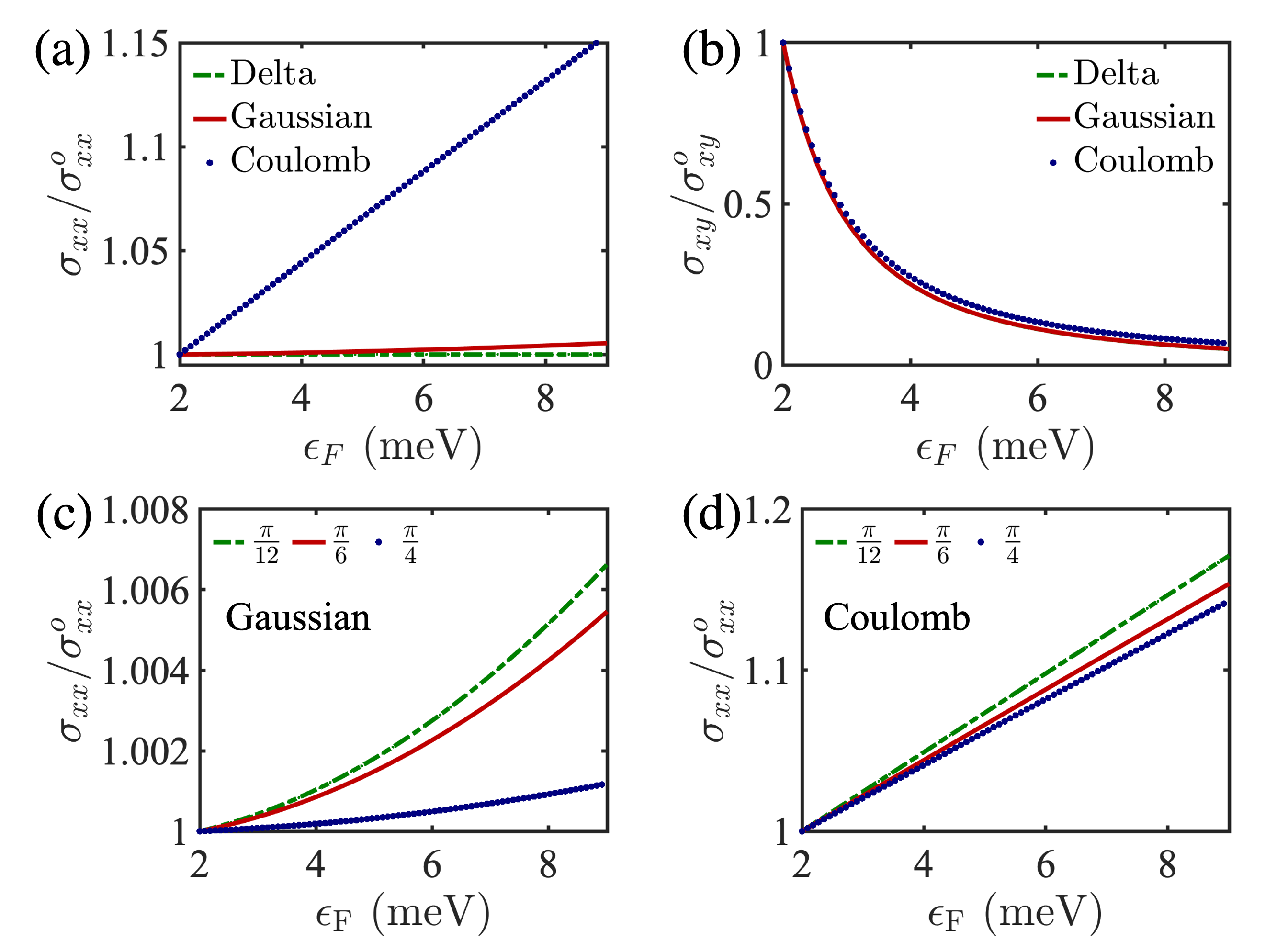}
    \caption{Conductivity for the $\alpha-T_3$ model when $m=0$. (a) $\sigma_{xx}$, and (b) $\sigma_{xy}$ for different disorder types. Comparison of $\sigma_{xx}$ for different values of $\psi$ is presented in (c) for Gaussian impurities, and in (d) for Coulomb impurities. $\sigma^o_{\alpha\beta}$ represent the values of $\sigma_{\alpha\beta}$ at $\epsilon_\mathrm{F}=2$meV. We chose $k_\mathrm{TF}=0.1$nm, $\sigma=6$nm, $b=0.01T$.}
    \label{fig:cond_mass_zero_no_omm}
\end{figure}
Before proceeding further, we briefly comment on the validity of the Boltzmann approximation assumed throughout this manuscript. The Boltzmann formalism is in general valid if the number of occupied Landau levels is large, i.e., when  the magnetic field $b\ll (\epsilon_\mathrm{F}-m)^2/e\hbar v^2$. For a field of $b\sim0.01T$ and for $(\epsilon_\mathrm{F}-m)\sim 4$meV, the ratio $e b\hbar v^2/(\epsilon_\mathrm{F}-m)^2\sim 0.4$, placing a rough limit on the range of validity of our formalism. For higher values of magnetic fields, the lower limit on $\epsilon_\mathrm{F}$ increases as well. Nonetheless, we plot the respective quantities over a wide range of the Fermi energy to demonstrate the qualitative trend and the transition from the valence to the conduction bands.

We first discuss the case of massless fermions. Even though the energy spectrum is independent of the parameter $\psi$, the eigenfunctions are $\psi$-dependent, and therefore one may expect qualitative and/or quantitative differences in the transport behavior. In particular, we find
\begin{align}
|\langle\mathbf{k}^{\prime}| \mathbf{k}\rangle|^2&= \frac{1}{4} \cos ^2\left(\frac{\phi -\phi'}{2}\right) (\cos (\phi -\phi')+3)\nonumber\\
& + \frac{1}{8}\cos (4 \psi ) \sin ^2(\phi -\phi'),
\end{align}
that depends non-trivially on $\psi$. Thus the scattering rate and conductivity depend crucially on $\psi$ as well. We evaluate the conductivity by numerically calculating conductivity from Eq.~\ref{Eq_sigmas}. In Fig.~\ref{fig:cond_mass_zero_no_omm} we plot the conductivity for the massless $\alpha-T_3$ model. We find that for $\delta-$disorder, the longitudinal conductivity $\sigma_{xx}$ is independent of the Fermi energy, similar to graphene. This is because for a $\delta-$type disorder, the scattering rate $w_{\mathbf{k}\mathbf{k}'}$ is momentum independent, and therefore on increasing the carrier density, the scattering time $\tau_\mathbf{k}$ decreases due to increase in the Fermi surface area. This decrease in the scattering time is exactly compensated by the increase in the density of states, resulting in a density independent conductivity. For Gaussian disorder, the scattering rate ($w_{\mathbf{k}\mathbf{k}'}\sim e^{-|\mathbf{k}-\mathbf{k}'|^2}$), decreases with increasing Fermi surface, and so is the case with Coulomb disorder ($w_{\mathbf{k}\mathbf{k}'}\sim 1/{|\mathbf{k}-\mathbf{k}'|^2}$). This accounts for the increased conductivity with increase in the Fermi energy, again resembling graphene. On the other hand, the Hall conductivity depends on the curvature of the Fermi surface, which is inversely proportional to the Fermi energy. The Hall conductivity is thus seen to decrease with increasing Fermi energy for all disorder profiles. Furthermore, we also compare the the effect of disorder for different values of the parameter $\psi$. We find that changing $\psi$ quantitatively affects the dependence on the Fermi energy, as shown in Figs.~\ref{fig:cond_mass_zero_no_omm}(c) and Figs.~\ref{fig:cond_mass_zero_no_omm}(d). Increasing $\psi$ from zero to $\pi/4$ decreases the magnitude of the conductivity. 

\begin{figure}
    \centering
    \includegraphics[width=\columnwidth]{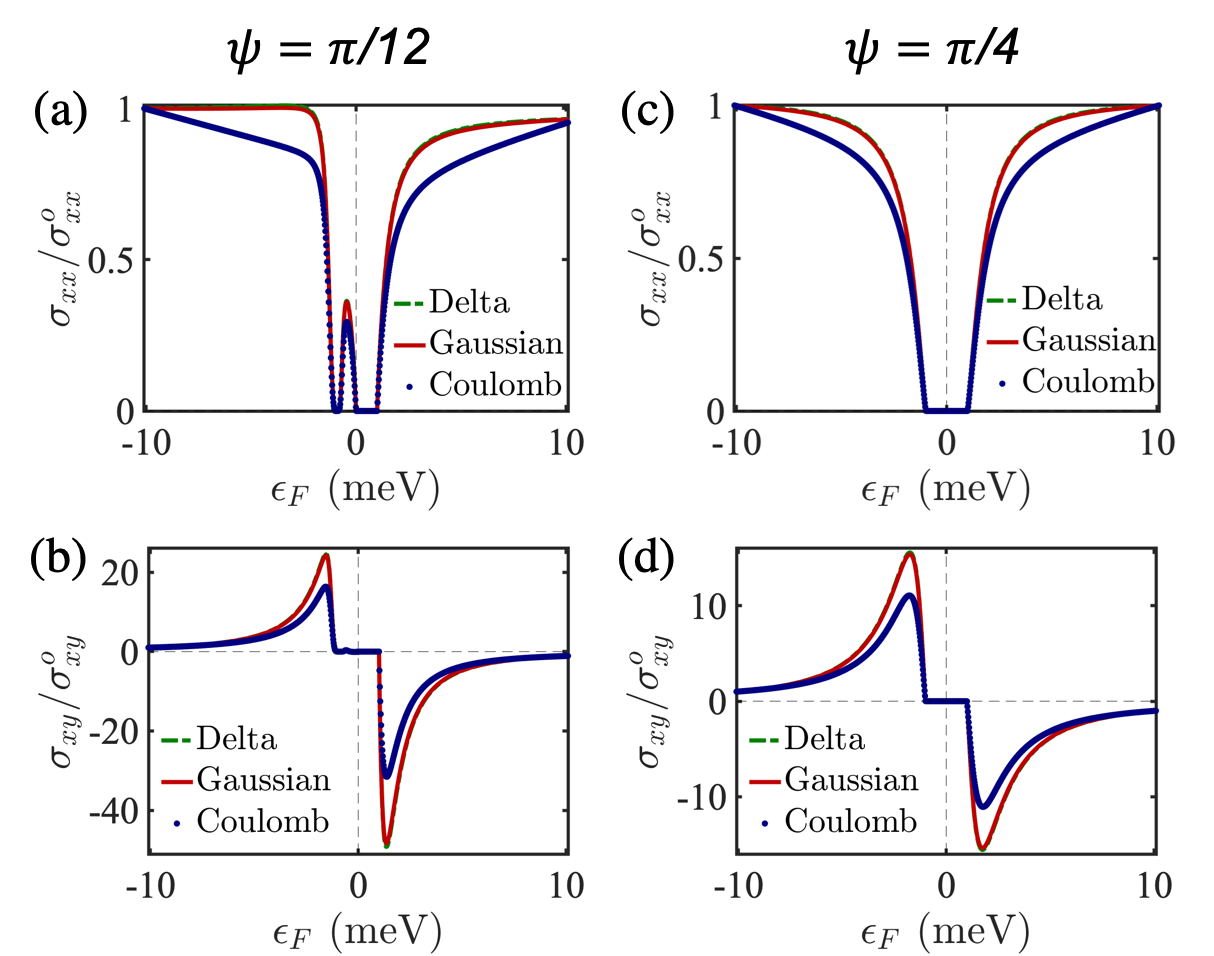}
    \caption{Conductivity in the case of $\alpha-T_3$ lattice with a finite mass term. Longitudinal conductivity for (a) $\psi=\pi/12$ and (c) $\psi=\pi/4$. Hall conductivity for (b) $\psi=\pi/12$ and (d) $\psi=\pi/4$. We chose $m=1$meV. All other parameters are chosen as in Fig.~\ref{fig:cond_mass_zero_no_omm}.}
    \label{fig:sigma_mass_1mev_del_zero}
\end{figure}
We next consider the presence of a finite nonzero mass. The bandstructure, as observed in Fig.~\ref{fig:bs_no_omm} acquires a gap between the lower band and the middle band and also between the middle band and the upper band. Furthermore, the electron-hole symmetry is typically violated, except when $\psi=\pi/4$. When $\psi=\pi/4$, the middle band remains dispersionless but is otherwise dispersive for any general value of $\psi$. Valley symmetry is, however, retained, and hence we focus only near one of the two valleys. In Fig.~\ref{fig:sigma_mass_1mev_del_zero} we plot the longitudinal as well as the Hall conductivity for two values of the parameter $\psi$. When $\psi=\pi/12$, and if the Fermi energy lies either in the valence or the conduction band, we observe the longitudinal conductivity (Fig.~\ref{fig:sigma_mass_1mev_del_zero}(a)) to be almost independent of electron density for Gaussian and $\delta-$impurities, indicating the reminiscence of Dirac-like dispersion in the limit $\psi\rightarrow 0$. The small peak in conductivity when $\epsilon_F<0$ is due to the dispersive middle band. For Coulomb impurities, we observe a linear dependence on Fermi energy similar to the massless case. When $\psi=\pi/4$, the density dependence is no longer constant for Gaussian and $\delta-$impurities indicating departure from the Dirac limit (Fig.~\ref{fig:sigma_mass_1mev_del_zero} (c)). The central peak is absent in this case due to the dispersionless middle band. The Hall conductivity peaks (with opposite signs) when the Fermi energy is just above or below the bandgap (Fig.~\ref{fig:sigma_mass_1mev_del_zero}(b) and Fig.~\ref{fig:sigma_mass_1mev_del_zero}(d)). The electron-hole asymmetry is also reflected in the Hall conductivity in Fig.~\ref{fig:sigma_mass_1mev_del_zero}(b). 

\begin{figure}
    \centering
    \includegraphics[width=\columnwidth]{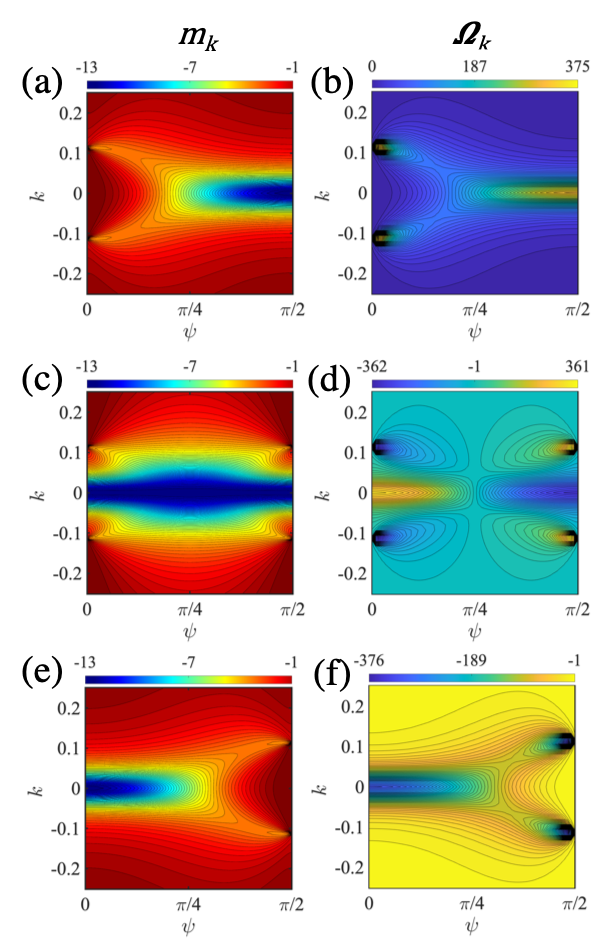}
    \caption{Orbital magnetic moment $m_\mathbf{k}$ for (a) the lower band, (c), the middle band, and (e) the upper band. as a function of the radial momentum and the parameter $\psi$ for the $\alpha-T_3$ lattice model. The corresponding Berry curvature is displayed on the right panels, (b), (d), and (f), respectively. 
    Here, $k$ is in the units of $1/a$, $m_k$ is in the units of $ave$, and $\Omega_k$ is in the units of $1/a^2$. The mass parameter was chosen $m=1$meV, and $\tau=+1$.}
    \label{fig:fig_omm_bc_mass1mev}
\end{figure}
\begin{figure}
    \centering
    \includegraphics[width=\columnwidth]{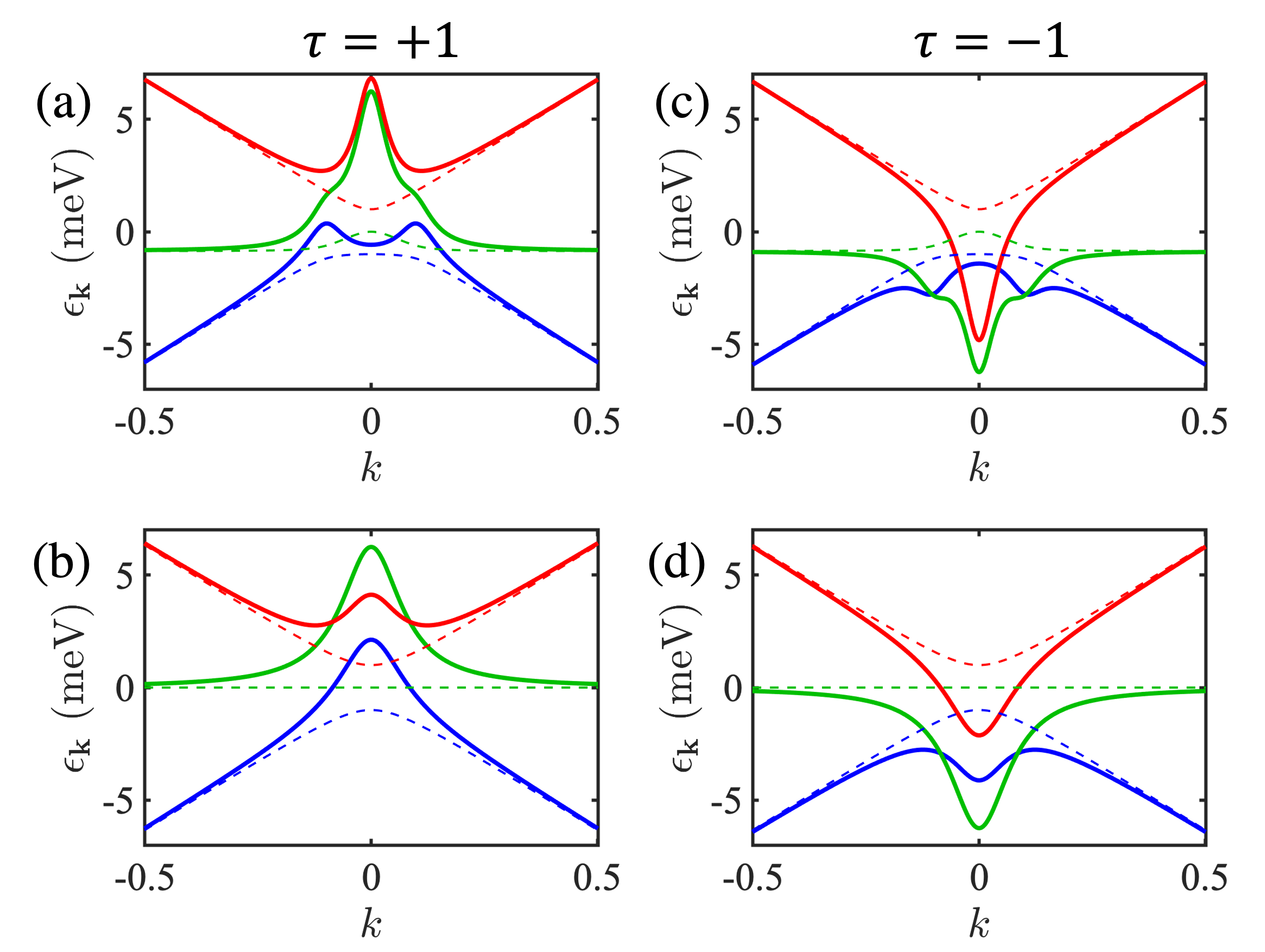}
    \caption{Bandstructure of the $\alpha-T_3$ model considering the effect of coupling to magnetic field due to a finite anomalous orbital magnetic moment . (a) $\tau=+1$, $\psi=\pi/12$, (a) $\tau=-1$, $\psi=\pi/12$, (a) $\tau=+1$, $\psi=\pi/4$,(a) $\tau=-1$, $\psi=\pi/4$. The dashed lines represent the dispersion without considering orbital magnetic moment. We choose $b=0.01$T.}
    \label{fig_bs_with_omm}
\end{figure}
\section{Berry phase effects in pseudospin-1 fermions}
\subsection{Orbital magnetic moment and Berry curvature}
Until now, we have neglected the effects of Berry curvature ($\boldsymbol{\Omega}_\mathbf{k}$) and the orbital magnetic moment ($\mathbf{m}_\mathbf{k}$). In the limit when $m=0$, both of these quantities vanish exactly. When the mass is nonzero, the orbital magnetic moment as well as the Berry curvature are finite and peak at certain points in the Brillouin zone. More generally, the presence of Berry curvature in a system implies breaking either time-reversal (TR) or spatial inversion (SI) symmetry. The Berry curvature is evaluated as:
\begin{align}
    \Omega^{n\gamma}_\mathbf{k} =i \sum_{n' \neq n} \frac{\langle n |d{H}/dk^\alpha| n' \rangle \langle n' |d{H}/dk^\beta|  n \rangle - (\alpha\leftrightarrow\beta)}{(\epsilon^n_\mathbf{k} - \epsilon^{n'}_\mathbf{k})^2},
\end{align}
where $n$ is the band index,  $\alpha$, $\beta$, and $\gamma$ represent the components of the vector, and $|n\rangle$ is the energy eigneket. 
Simultaneously, the self-rotation of the Bloch wave-packet also generates an intrinsic anomalous orbital magnetic moment that is given by~\cite{xiao2010berry} 
\begin{align}
        \mathbf{m}^n_{\mathbf{k}}&=\frac{-i e}{2\hbar}\bra{\nabla_{\mathbf{k}} n }\mathbf{\times}[{H}-\epsilon^n_\mathbf{k}]\ket{\nabla_\mathbf{k} n}.
\end{align}
In Fig.~\ref{fig:fig_omm_bc_mass1mev} we plot the orbital magnetic moment and the Berry curvature for all the three bands as a function of the radial momentum $k$ and the lattice parameter $\psi$. Specifically we plot the $z-$component of $\mathbf{m}_\mathbf{k}$ and $\boldsymbol{\Omega}_\mathbf{k}$ since the $x$ and $y$ components are zero. As discussed, both the orbital magnetic moment and the Berry curvature acquire finite values in the Brillouin zone. When $\psi\rightarrow 0$, the lower two bands touch each other away from $k=0$ (Fig.~\ref{fig:bs_no_omm} (b)), which generates a peak in the Berry curvature and the orbital magnetic moment. In the limit $\psi\rightarrow \pi/2$, the upper two bands touch each other away from $k=0$, which generates similar peaks as well. This feature is consistent with the fact that the model has a $\alpha\rightarrow1/\alpha$ duality, i.e., in the limit of $\psi\rightarrow 0$ and $\psi\rightarrow \pi/2$ the model maps onto the graphene lattice. We restrict our discussion to $\tau=+1$ as switching the valley index only reverses the sign of the orbital magnetic moment and the Berry curvature. Importantly, we also note the qualitative differences in the Berry curvature distribution and the orbital magnetic moment between the spin-1/2 Dirac fermions in a gapped graphene model~\cite{xiao2007valley} and the pseudospin-1 fermions here. In the model of gapped graphene, these quantities peak exactly at the $\mathbf{K}$-point unlike what we observe in Fig.~\ref{fig:fig_omm_bc_mass1mev}.

A nonzero flux of Berry curvature in the Brillouin zone leads to  anomalous Hall conductivity, but since here the overall Berry curvature vanishes on adding contributions from both the valleys, anomalous Hall conductivity is zero. However, as we shall shortly see, a small but finite magnetic field results in a finite anomalous Hall conductivity (along with the usual Lorentz-force driven Hall conductivity) due to breakdown of valley symmetry.
More specifically, the anomalous magnetic moment couples to the applied magnetic field as $\epsilon^n_\mathbf{k}\rightarrow \epsilon^n_\mathbf{k} - \mathbf{m}^n_\mathbf{k}\cdot\mathbf{b}$, where $\mathbf{b}$ is the applied magnetic field. Since the sign of the orbital magnetic moment is opposite at both the valleys, the valley symmetry is lost on application of magnetic field. This is highlighted in Fig.~\ref{fig:fig_omm_bc_mass1mev}, which plots the energy dispersion of $\alpha-T_3$ model both in the presence and absence of orbital magnetic moment. Interestingly, we find that the energy dispersion not only shifts upwards or downwards depending on the valley, multiple Fermi surfaces may result depending on the value of the chemical potential due to unequal shifting of various bands. For example, in Fig.~\ref{fig_bs_with_omm} (a) and Fig.~\ref{fig_bs_with_omm} (b), if the Fermi level intersects in the upper band, one may encounter a situation where on application of a magnetic field, an electron-like Fermi surface additionally acquires one or two hole-like Fermi surfaces due to the bending of both the upper band and the middle band. Similarly, if the chemical potential intersects the lower band, then an analogous situation may arise in the other valley (Fig.~\ref{fig_bs_with_omm} (c) and Fig.~\ref{fig_bs_with_omm} (d)). Furthermore, an interesting situation is encountered if the chemical potential intersects close by the middle band, where the Fermi surface consists of electron and hole-like pockets on the positive and negative valleys, respectively, or vice-versa, depending on the direction of the magnetic field. This may result in exact compensation of charges for $\psi=\pi/4$ at suitable value of the chemical potential (Fig.~\ref{fig_bs_with_omm} (b) and Fig.~\ref{fig_bs_with_omm} (d)), yielding a compensated semimetal. We next discuss the implications of the effect of orbital magnetic moment and Berry phase on electronic transport. 

\begin{figure}
    \centering
    \includegraphics[width=\columnwidth]{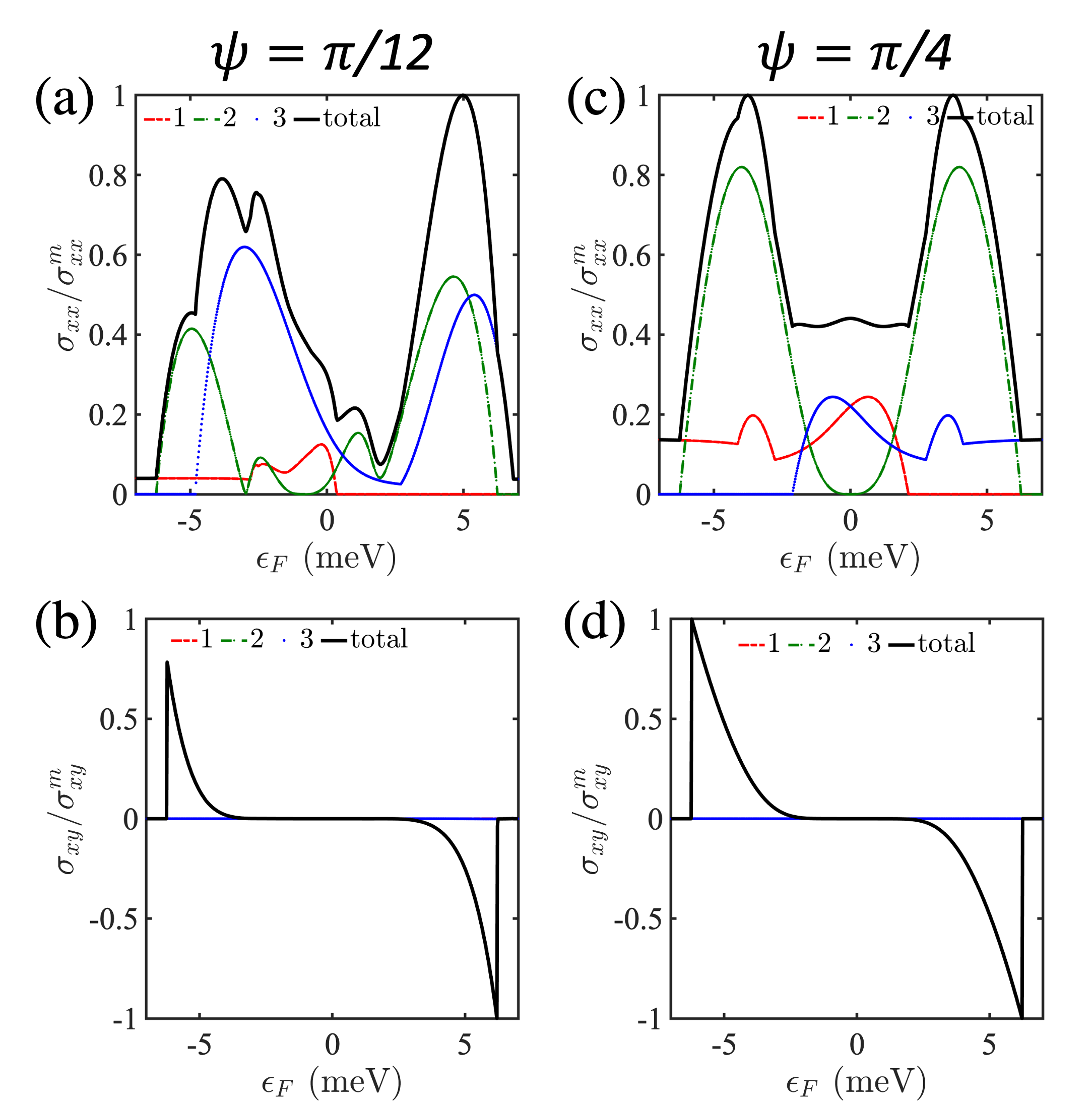}
    \caption{(a) Longitudinal conductivity as a function of the chemical potential including the effect of orbital magnetic moment. The red, green, and blue curves, indicate the contribution from the lower (1), middle (2), and upper (3) bands respectively. The black curve indicates the net contribution. Here $m=1$meV, and $b=0.01$T. (b) The corresponding Hall conductivity. The dominant contribution is from the middle band. All plots are normalized with respect to their absolute maximum values. }
    \label{fig:sig_del_one_1}
\end{figure}

\begin{figure}
    \centering
    \includegraphics[width=\columnwidth]{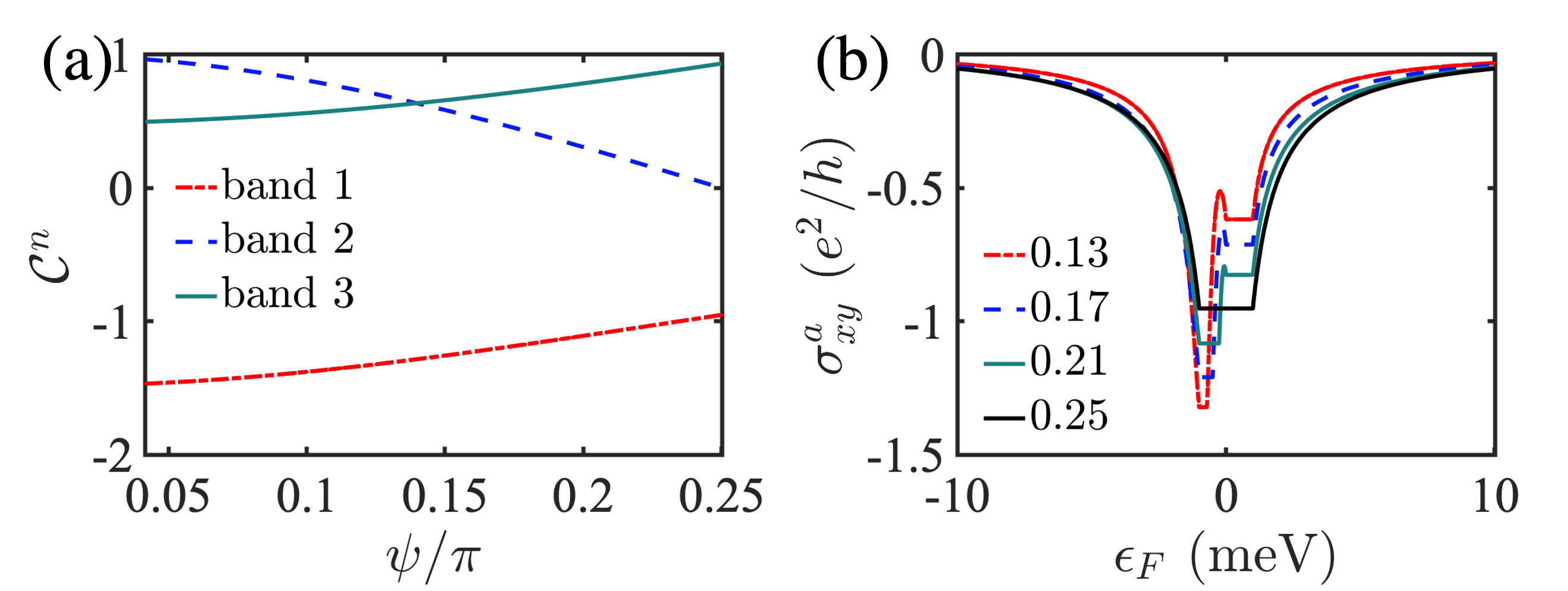}
    \caption{(a) Chern number at $\tau=+1$ as a function of the parameter $\psi$. (b) The anomalous Hall conductance contribution of $\tau=+1$ as a function of the Fermi energy. We chose $m=1$meV. Since $\mathcal{C}^n$ at $\tau=-1$ is negative of the value at $\mathcal{C}^n$, the total anomalous Hall conductance given by the contribution of the two valleys is zero. }
    \label{fig:chern_anomsxy}
\end{figure}

\begin{figure}
    \centering
    \includegraphics[width=\columnwidth]{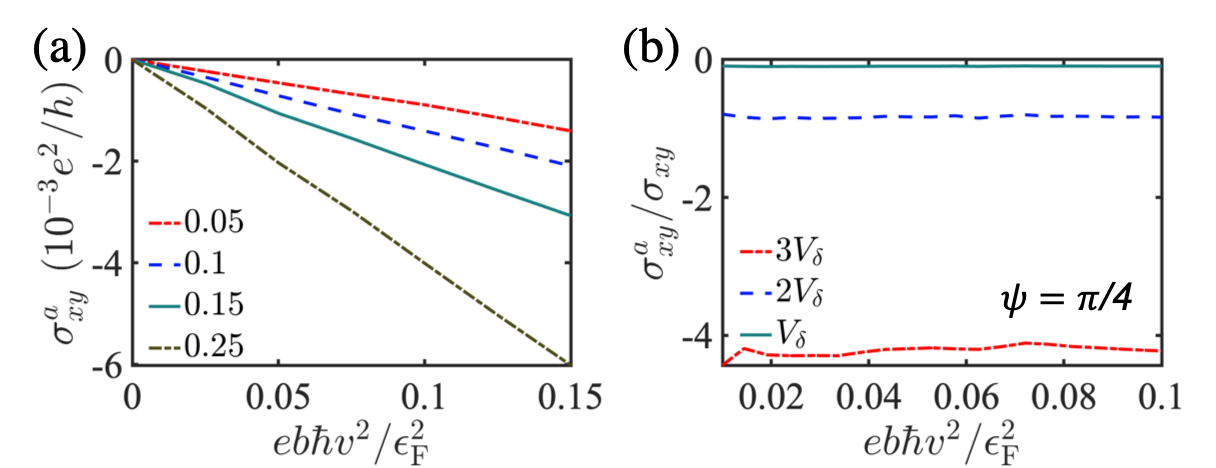}
    \caption{(a) The total anomalous Hall conductivity $\sigma_{xy}^a$ as a function of the magnetic field. (b) The ratio of the anomalous ($\sigma_{xy}^a$) and the conventional Hall conductivity ($\sigma_{xy}$) for three different values of the disorder potential $V_\delta$, and fixed $\psi=\pi/4$.}
    \label{fig:sigma_anom_vs_b}
\end{figure}
\subsection{Unconventional transport}
In Fig.~\ref{fig:sigma_mass_1mev_del_zero}, we plot the longitudinal conductivity for the $\alpha-T_3$ lattice in the presence of orbital magnetic coupling as a function of the Fermi energy. The behaviour of longitudinal conductivity is strikingly different when compared to the case when the orbital magnetic moment is absent. In Fig.~\ref{fig:sigma_mass_1mev_del_zero} (a), we observe that the lower band contributes only just below the Fermi energy, and the middle band has two large peaks far above and below $\epsilon_\mathrm{F}$, each accompanied by two smaller peaks. The upper band has a contribution above and below $\epsilon_\mathrm{F}$ as well. As expected, the conductivity profile is congruent with the bandstructure presented in Fig.~\ref{fig_bs_with_omm} (a) and Fig.~\ref{fig_bs_with_omm} (b). The Hall conductivity (Fig,~\ref{fig:sig_del_one_1} (c)), being sensitive to the Fermi surface curvature, has a primary contribution from the middle band that peaks when the band is either close to being fully occupied or fully empty. We find that the Hall contribution from the upper and the lower band are at least one order lesser than the contribution of the middle band. Fig.~\ref{fig:sig_del_one_1} (b) and Fig.~\ref{fig:sig_del_one_1} (d) present the longitudinal and Hall conductivity when $\psi=\pi/4$, where electron-hole symmetry is restored in this case. Furthermore, we find that unlike Fig.~\ref{fig:sigma_mass_1mev_del_zero} and Fig.~\ref{fig:cond_mass_zero_no_omm}, the qualitative trend is very less sensitive to the nature of the underlying impurities, because the primary deviations on account of the orbital magnetic moment further remove the \textit{Diracness} in the dispersion of the quasiparticles.

\begin{figure}
    \centering
    \includegraphics[width=\columnwidth]{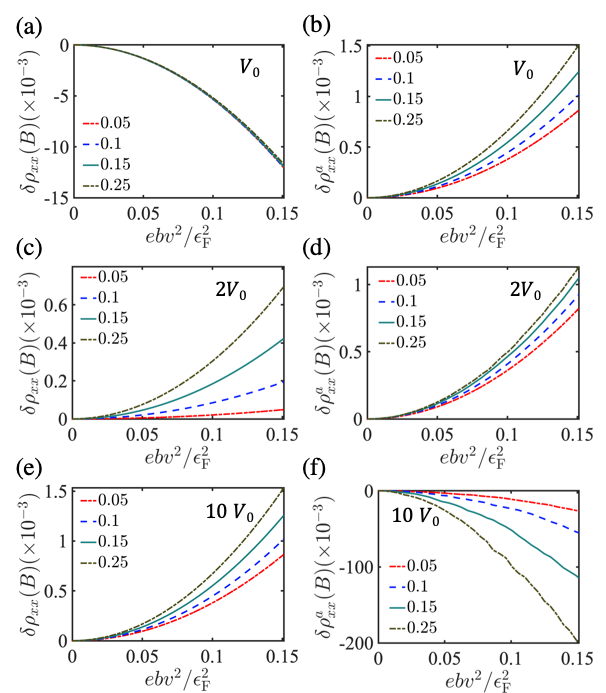}
    \caption{Magnetoresistivity in the $\alpha-\mathcal{T}_3$ model for various values of the parameter $\psi$. (a), (c), and (e), plot the contribution due to conventional Hall effect, for different disorder strengths; (b), (d), and (f) plot the contribution due to anomalous Hall effect.}
    \label{fig:magres1}
\end{figure}

The non-trivial Berry curvature of the energy bands guarantees a nonzero Chern number that is given by 
\begin{align}
    \mathcal{C}^n = \frac{1}{2\pi} \int{d^2 \mathbf{k} \hspace{1mm} \boldsymbol{\Omega}_\mathbf{k}^n}.
\end{align}
In Fig.~\ref{fig:chern_anomsxy} (a), we plot the Chern number for the massive $\alpha-\mathcal{T}_3$ lattice at the  $\tau=+1$ valley as a function of the parameter $\psi$. The presence of nonzero Chern number gives rise to anomalous Hall conductivity, given by 
\begin{align}
    \sigma_{xy}^a = \sum\limits_n\int{\frac{d^2\mathbf{k}}{(2\pi)^2} \hspace{1mm} \boldsymbol{\Omega}_\mathbf{k}^n f^n_0(\mathbf{k})}.
\end{align}
In Fig.~\ref{fig:chern_anomsxy} (b), we plot the anomalous Hall conductance contribution of $\tau=+1$ as a function of the Fermi energy. Since $\mathcal{C}^n$ at $\tau=-1$ is of the opposite sign of $\mathcal{C}^n$ at $\tau=+1$, the total anomalous Hall conductance given by the contribution of the two valleys is zero. 

As noted before, the presence of a magnetic field induces a Zeeman-like contribution to the energy spectrum that is of opposite sign at both valleys; hence the contribution of the anomalous Hall conductivity at both valleys does not cancel out due to their Fermi surfaces being dissimilar as a result of coupling to the magnetic field. This yields a finite anomalous Hall conductance, as plotted in Fig.~\ref{fig:sigma_anom_vs_b} (a), which shows $\sigma_{xy}^a$ as a function of $b$ for various values of the parameter $\psi$. In evaluating $\sigma_{xy}^a$, we have fixed the Fermi energy so that it intersects only the upper band with a single Fermi surface, and thus contributions from the middle and the lower bands are zero. In Fig.~\ref{fig:sigma_anom_vs_b} (b) we plot the ratio of the anomalous Hall conductivity ($\sigma_{xy}^a$) to the conventional conductivity ($\sigma_{xy}$) for three different values of the disorder potential. While the strength of the disorder potential doesn't affect the anomalous Hall conductivity, it does so, the conventional Hall conductivity. Depending on the strength of the disorder, $\sigma_{xy}^a/\sigma_{xy}$ can be either lesser or greater than one. The ratio, however, is largely independent of the magnetic field since both the anomalous and the conventional Hall conductivity increase linearly with $b$. Last, we also highlight another important difference between conventional and anomalous Hall conductivity: their opposite signs. 

\subsection{Magnetoresistance}
We now discuss the (longitudinal) magnetoresitance in the $\alpha-\mathcal{T}_3$ model. 
Magnetoresistance from a single parabolic band is ideally zero (but conductivity is typically nonzero). Deviations from parabolicity and/or multi-band effects may however yield a finite magnetoresistance. In the present case, the dissimilarity of the Fermi surfaces on account of the magnetic field yields a finite longitudinal magnetoconductivity $\sigma_{xx}(B)$. Depending on the how the Hall conductivity varies with the magnetic field (and how much), we anticipate a non-trivial magnetoresistivity in the $\alpha-\mathcal{T}_3$  model. Before explicitly evaluating the magnetoresistance in the $\alpha-\mathcal{T}_3$ model, we consider a  generic toy model for magnetoresistance. The Hall conductivity, whether anomalous or conventional, typically has a linear dependence on the magnetic field, while the longitudinal magnetoconductivity is typically quadratic. This motivates us to consider the following conductivity tensor:
\begin{align}
    \hat{\sigma}=\begin{pmatrix}
    \sigma^o+\beta h^2 & \gamma h\\
    -\gamma h & \sigma^o+\beta h^2
    \end{pmatrix},
\end{align}
where $h$ is the dimensionless magnetic field, the dimensionless longitudinal conductivity at zero magnetic field is $\sigma^o$, $\beta$ is the quadratic coefficient of the longitudinal conductivity, and $\gamma$ is the linear coefficient of the Hall conductivity. The magnetoresistivity $\delta\rho(B)$ is defined as: 
\begin{align}
    \delta\rho(B) = \frac{\rho_{xx}(B) - \rho_{xx}(0)}{\rho_{xx}(0)},
\end{align}
where $\rho_{xx}(B)$ is the (1,1) component of the resistivity tensor $\hat{\rho}=\hat{\sigma}^{-1}$. In the current model, 
\begin{align}
    \delta\rho(B) = \frac{\sigma^o  \left(\beta  h^2+\sigma^o \right)}{\beta
   ^2 h^4+h^2 \left(2 \beta  \sigma^o +\gamma
   ^2\right)+(\sigma^o) ^2}-1.
\end{align}
Clearly, at lower values of magnetic field, the magnetoresistivity is positive if $\beta<-\gamma^2/\sigma^o$; for all other cases the magnetoresistivity is negative. The sign of magnetoresistivity is independent of the sign of $\gamma$, or in other words independent of the type of charge carriers in the system, as expected. In the low-field limit, when the magnetoconductivity is negative and if quadratic coefficient $\beta$ is negative and below $<-\gamma^2/\sigma^o$, we obtain positive magnetoresistivity. Above a critical value of the magnetic field, $h_c=(\sqrt{-\gamma^2-\beta\sigma^o})/|\beta|$, the magnetoresistivity eventually becomes negative. 

For conventional Hall conductivity, the coefficient $\gamma$ depends on the disorder strength, but $\gamma$ is independent of disorder for anomalous Hall conductivity. In the present case, the contribution to the coefficient $\beta$ mainly arises from the coupling of the anomalous orbital magnetic moment to the magnetic field, and is therefore independent of the disorder strength. Finally, $\sigma^o$ clearly is dependent on the strength of the disorder. It is thus likely that the magnetoresistivity contribution from the conventional Hall effect or the anomalous Hall effect can be either positive or negative due to competing effects of $\beta$ (independent of disorder) and $\gamma$ (dependent on disorder), and $\sigma^o$ (dependent on disorder). Although, our model is applicable to the system considered in Ref.~\cite{zhou2019valley} (gapped-graphene), a discussion on the possibility of both signs of magnetoresistance remains absent in their work. We conjecture that as a result of orbital magnetic moment coupling, both positive and negative magnetoresistance should be observed in gapped-graphene (Ref.~\cite{zhou2019valley}) as well. 

In Fig.~\ref{fig:magres1} we plot the magnetoresistivity for the $\alpha-\mathcal{T}_3$ model, separately for the conventional and the anomalous Hall contributions, thus revealing their qualitative differences. As the strength of disorder is increased, we observe that the conventional contribution to magnetoresistivity switches sign from negative to positive, and decreases in the overall magnitude, because on increasing the disorder strength, the condition $\beta<-\gamma^2/\sigma^o$ is satisfied much more easily. Even though, $1/\sigma^o$ increases on increasing  disorder, $\gamma^2$ decreases even more rapidly, thus making the requirement for $\beta$ to be lesser than $-\gamma^2/\sigma^o$ easier to be realized; yielding positive magnetoresistivity. Contrary to this, we find that on increasing disorder, the anomalous contribution to the magnetoresistivity changes sign from positive to negative. This is because $1/\sigma^o$ increases on increasing  disorder, but $\gamma^2$ remains constant, therefore making it harder to realize the condition $\beta<-\gamma^2/\sigma^o$ for positive magnetoresistivity. Furthermore, the magnitude of magnetoresistivity increases as well, because the magnitude of the anomalous Hall conductivity remains a constant, while that of $\sigma^o$ decreases. We close this section with the remark that the anomalous electronic transport quantities will also reflect on the anomalous thermoelectric quantities. For instance, the derivative of the Hall conductivity with respect to the Fermi energy yields the Nernst effect, which should also be substantially measurable here. We defer the evaluation of anomalous thermoelectric coefficients in the $\alpha-\mathcal{T}_3$ model to future works. 

\section{Conclusions} While the $\alpha-\mathcal{T}_3$ model of pseudospin-1 fermions owes similarities to the model of pseudospin-1/2 Dirac fermions, the effects on electronic transport are more nuanced. Even though their energy dispersion is qualitatively similar, presence of the middle band in the $\alpha-\mathcal{T}_3$ lattice changes the wavefunctions' topology and, more so, the distribution of the Berry curvature and the orbital magnetic moment in the Brillouin zone. Here we studied electronic transport in the $\alpha-\mathcal{T}_3$ model and evaluate their conventional as well as anomalous transport responses in the presence of weak electric and magnetic fields. First, we listed out the differences arising in the conventional conductivity (longitudinal and Hall) for three different types of impurities: $\delta$-correlated disorder, Gaussian disorder, and Coulomb disorder. Second, we studied the Berry curvature and orbital magnetic moment of the pseudospin-1 fermions and evaluate their anomalous transport responses. An important finding of this work is that coupling the orbital magnetic moment to the external magnetic field breaks valley symmetry, results in finite and measurable corrections to the conductivity, and yields anomalous Hall conductivity due to the nonzero distribution of the flux of the Berry curvature. Third, we found that a finite longitudinal magnetoresistance is induced by the orbital magnetic moment, which can be of either sign: positive or negative, depending on the amount of disorder, a feature which can be contrasted to pseudospin-1/2 fermions discussed in Ref.~\cite{zhou2019valley}. Recent material advances and upcoming experiments on cold atoms and Hg$_{1-x}$Cd$_x$Te quantum wells that may realize pseudospin-1 fermions makes our study even more apposite. 

\bibliography{biblio.bib}
\end{document}